\documentclass[aps,prl,twocolumn,groupedaddress,final,floatfix]{revtex4}
\usepackage{amsmath,amssymb,amsfonts}
\usepackage{bm}
\usepackage{bbm}
\usepackage{slashed}
\usepackage{rotating}
\usepackage[vcentermath]{youngtab}

\begin{document}

\newcommand{\alps}{\ensuremath{\alpha_s}}
\newcommand{\qbar}{\bar{q}}
\newcommand{\beq}{\begin{equation}}
\newcommand{\eeq}{\end{equation}}
\newcommand{\beqa}{\begin{eqnarray}}
\newcommand{\eeqa}{\end{eqnarray}}
\newcommand{\gs}{g_{\pi NN}}
\newcommand{\gw}{f_\pi}
\newcommand{\mq}{m_Q}
\newcommand{\mn}{m_N}
\newcommand{\bb}{\langle}
\newcommand{\kb}{\rangle}
\newcommand{\st}{\ensuremath{\sqrt{\sigma}}}
\newcommand{\rvec}{\mathbf{r}}
\newcommand{\bvec}[1]{\ensuremath{\mathbf{#1}}}
\newcommand{\bra}[1]{\ensuremath{\bb#1|}}
\newcommand{\ket}[1]{\ensuremath{|#1\kb}}
\newcommand{\gft}{\ensuremath{\gamma_{FT}}}
\newcommand{\bfalp}{\mbox{\boldmath{$\alpha$}}}
\newcommand{\bfnab}{\mbox{\boldmath{$\nabla$}}}
\newcommand{\bfpi}{\mbox{\boldmath{$\pi$}}}
\newcommand{\bfsig}{\mbox{\boldmath{$\sigma$}}}
\newcommand{\bftau}{\mbox{\boldmath{$\tau$}}}
\newcommand{\bfrho}{\ensuremath{\bm{\rho}}}
\newcommand{\bflambda}{\ensuremath{\bm{\lambda}}}
\newcommand{\bfchi}{\ensuremath{\bm{\chi}}}
\newcommand{\bfR}{\ensuremath{\bvec{R}}}
\newcommand{\spup}{\uparrow}
\newcommand{\spd}{\downarrow}
\newcommand{\hbarom}{\frac{\hbar^2}{m_Q}}
\newcommand{\half}{\ensuremath{\frac{1}{2}}}
\newcommand{\shalf}{{\scriptstyle\frac{1}{2}}}
\newcommand{\third}{{\frac{1}{3}}}
\newcommand{\sthird}{{\scriptstyle\frac{1}{3}}}
\newcommand{\vnn}{\ensuremath{\hat{v}_{NN}}}
\newcommand{\argonne}{\ensuremath{v_{18}}}
\newcommand{\lqcd}{\ensuremath{\mathcal{L}_{QCD}}}
\newcommand{\lgf}{\ensuremath{\mathcal{L}_g}}
\newcommand{\lqm}{\ensuremath{\mathcal{L}_q}}
\newcommand{\lqg}{\ensuremath{\mathcal{L}_{qg}}}
\newcommand{\nn}{\ensuremath{NN}}
\newcommand{\hpnd}{\ensuremath{H_{\pi N\Delta}}}
\newcommand{\hpqq}{\ensuremath{H_{\pi qq}}}
\newcommand{\fpnn}{\ensuremath{f_{\pi NN}}}
\newcommand{\fpnd}{\ensuremath{f_{\pi N\Delta}}}
\newcommand{\fpqq}{\ensuremath{f_{\pi qq}}}
\newcommand{\ylm}{\ensuremath{Y_\ell^m}}
\newcommand{\ylmc}{\ensuremath{Y_\ell^{m*}}}
\newcommand{\qbh}{\hat{\bvec{q}}}
\newcommand{\xbh}{\hat{\bvec{X}}}
\newcommand{\dt}{\Delta\tau}
\newcommand{\qmag}{|\bvec{q}|}
\newcommand{\pmag}{|\bvec{p}|}
\newcommand{\oas}{\ensuremath{\mathcal{O}(\alpha_s)}}
\newcommand{\vtxb}{\ensuremath{\Lambda_\mu(p',p)}}
\newcommand{\vtxp}{\ensuremath{\Lambda^\mu(p',p)}}
\newcommand{\pwqp}{e^{i\bvec{q}\cdot\bvec{r}}}
\newcommand{\pwqm}{e^{-i\bvec{q}\cdot\bvec{r}}}
\newcommand{\gsa}[1]{\ensuremath{\bb#1\kb_0}}
\newcommand{\exv}[1]{\ensuremath{\bb\hat{#1}\kb}}
\newcommand{\oer}[1]{\mathcal{O}\left(\frac{1}{\qmag^{#1}}\right)}
\newcommand{\nub}[1]{\overline{\nu^{#1}}}
\newcommand{\balph}{\mbox{\boldmath{$\alpha$}}}
\newcommand{\bgam}{\mbox{\boldmath{$\gamma$}}}
\newcommand{\epf}{E_\bvec{p}}
\newcommand{\epfp}{E_{\bvec{p}'}}
\newcommand{\eka}{E_{\alpha\kappa}}
\newcommand{\ekaq}{(E_{\alpha\kappa})^2}
\newcommand{\ekap}{E_{\alpha'\kappa}}
\newcommand{\ekpa}{E+{\alpha\kappa_+}}
\newcommand{\ekma}{E_{\alpha\kappa_-}}
\newcommand{\ekp}{E_{\kappa_+}}
\newcommand{\ekm}{E_{\kappa_-}}
\newcommand{\ekpap}{E_{\alpha'\kappa_+}}
\newcommand{\ekmap}{E_{\alpha'\kappa_-}}
\newcommand{\yjm}[1]{\mathcal{Y}_{jm}^{#1}}
\newcommand{\ysa}[3]{\mathcal{Y}_{#1,#2}^{#3}}
\newcommand{\ysc}{\tilde{y}}
\newcommand{\enm}{\varepsilon_{NM}}
\newcommand{\Scg}[6]
	{\ensuremath{S^{#1}_{#4}\:\vphantom{S}^{#2}_{#5}
 	 \:\vphantom{S}^{#3}_{#6}\,}}
\newcommand{\Kmat}[6]
	{\ensuremath{K\left[\begin{array}{ccc} 
	#1 & #2 & #3 \\ #4 & #5 & #6\end{array}\right]}}
\newcommand{\irt}{\ensuremath{\frac{1}{\sqrt{2}}}}
\newcommand{\irth}{\ensuremath{\frac{1}{\sqrt{3}}}}
\newcommand{\irs}{\ensuremath{\frac{1}{\sqrt{6}}}}
\newcommand{\rtoth}{\ensuremath{\sqrt{\frac{2}{3}}}}
\newcommand{\Tg}{\ensuremath{\mathsf{T}}}
\newcommand{\irrep}[1]{\ensuremath{\mathbf{#1}}}
\newcommand{\cirrep}[1]{\ensuremath{\overline{\mathbf{#1}}}}
\newcommand{\Fij}{\ensuremath{\hat{F}_{ij}}}
\newcommand{\Fqij}{\ensuremath{\hat{F}^{(qq)}_{ij}}}
\newcommand{\Fsij}{\ensuremath{\hat{F}^{(qs)}_{ij}}}
\newcommand{\Opij}{\mathcal{O}^p_{ij}}
\newcommand{\titj}{\bftau_i\cdot\bftau_j}
\newcommand{\sisj}{\bfsig_i\cdot\bfsig_j}
\newcommand{\tens}{S_{ij}}
\newcommand{\LS}{\bvec{L}_{ij}\cdot\bvec{S}_{ij}}
\newcommand{\TT}{\Tg_i\cdot\Tg_j}
\newcommand{\chet}{\ensuremath{\chi ET}}
\newcommand{\chpt}{\ensuremath{\chi PT}}
\newcommand{\chsy}{\ensuremath{\chi\mbox{symm}}}
\newcommand{\lchi}{\ensuremath{\Lambda_\chi}}
\newcommand{\lcon}{\ensuremath{\Lambda_{QCD}}}
\newcommand{\dcpsi}{\ensuremath{\bar{\psi}}}
\newcommand{\dc}[1]{\ensuremath{\overline{#1}}}
\newcommand{\llo}{\ensuremath{\mathcal{L}^{(0)}_{\chet}}}
\newcommand{\lchet}{\ensuremath{\mathcal{L}_{\chi}}}
\newcommand{\Dmu}{\ensuremath{\mathcal{D}_\mu}}
\newcommand{\Dsl}{\ensuremath{\slashed{\mathcal{D}}}}
\newcommand{\comm}[2]{\ensuremath{[#1,#2]}}
\newcommand{\acomm}[2]{\ensuremath{\{#1,#2\}}}

\title{Variational Monte Carlo study of pentaquark states}

\author{Mark W.\ Paris}
\email{mparis@jlab.org}
\affiliation{Thomas Jefferson National Accelerator Facility,
Theory Group, 12000 Jefferson Avenue MS12H2, Newport News, Virginia, 23606}

\date{\today}
\begin{abstract}
Accurate numerical solution of the five-body Schr\"{o}dinger
equation is effected via variational Monte Carlo. The spectrum is assumed 
to exhibit a narrow resonance with strangeness $S=+1$. A
fully antisymmetrized and pair-correlated five-quark
wave function is obtained for the assumed non-relativistic Hamiltonian
which has spin, isospin, and color dependent pair interactions and many-body
confining terms which are fixed by the non-exotic spectra.
Gauge field dynamics are modeled via flux tube exchange 
factors.  The energy determined for the ground states with 
$J^\pi=\shalf^-(\shalf^+)$ is 2.22 GeV (2.50 GeV).
A lower energy negative parity state is consistent with recent 
lattice results. The short-range structure of the state is analyzed 
via its diquark content.
\end{abstract}

\maketitle

A system of interacting, non-relativistic constituent quarks is the
most simple, realistic model of hadronic systems. Solving the many-body
Schr\"{o}dinger equation to determine wave 
functions within this simple model is still
a formidable task owing to the strong flavor, spin, and color dependence
of the quark-quark interaction and traditionally requires some array of 
approximate methods to solve it.
The controversial status of the recent evidence of a flavor exotic
five-quark state warrants a careful treatment of this strongly
interacting system. In this letter, the exercise of 
determining the wave function of five interacting constituent quarks in 
the flavor-exotic multiquark hadronic sector is solved using the 
variational Monte Carlo (VMC) technique. This technique is known
to yield upper bounds on the ground state energy accurate to the level
of a few percent in light nuclei with the number of nucleons $A\le 6$
\cite{Pudliner:1995wk}.

Recent experimental evidence \cite{Nakano:2003qx, Barth:2003es,
Stepanyan:2003qr, Kubarovsky:2003fi, Barmin:2003vv, Abdel-Bary:2004ts,
Asratyan:2003cb, Airapetian:2003ri, Chekanov:2004kn, Alt:2003vb,
Aslanyan:2004gs, Dzierba:2004db} has revived interest 
in the multiquark flavor-exotic sector of the hadronic spectrum.
Various model calculations have been 
reported \cite{Strottman:1979qu,Weinstein:1982gc,Praszalowicz:1987em,
Lipkin:1987sk, Carlson:1991zt, Diakonov:1997mm, Jaffe:2003sg,
Karliner:2003dt, Carlson:2003pn, Hiyama:2005cf} which study the 
existence of a strangeness $S=+1$ resonance, dubbed $\theta^+$, about 
100 MeV above threshold to nucleon-kaon decay with low mass of 1540(2) 
MeV and possibly extremely narrow width of $0.9(3)$ MeV
\cite{Eidelman:2004wy}. Lattice results are available, including Refs.
\cite{Mathur:2004jr,Lasscock:2005tt}.

The calculation of Ref.\cite{Diakonov:1997mm}, 
working in the chiral quark soliton model, of an 
extremely small width, $<15$ MeV stimulated 
experimental searches with some positive signals, although all of them
had very low statistics with at most $\sim 100$ counts above a comparable
background \cite{Dzierba:2004db}.
Preliminary high statistics photoproduction data of the reaction
$\gamma p \rightarrow n K_S$ from the CLAS collaboration \cite{diVita}
sets an upper limit on the yield of the $\theta^+$ relative 
to the $\Lambda^*(1520)$ yield at 0.2\%. No definitive structure in
the $nK$-invariant mass spectrum is observed at 1540 MeV in this
experiment. No experimental information is available on the spin or 
parity of the state in any experiment done to date.

\begin{figure}[b]
\includegraphics[ width=200pt, keepaspectratio, angle=0, clip ]{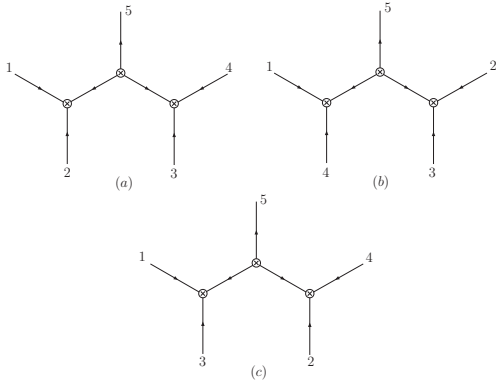}
\caption{\label{fig:FTtopo} Topological configurations of flux tubes.
In general, configurations of quarks do not lie in the plane.}
\end{figure}

We study the $\theta^+$ in the non-relativistic flux tube 
quark model with one-gluon exchange (OGE) and one-pion exchange (OPE) 
(between the light quarks). We take seriously the possibility that 
$\theta^+$ is a narrow resonance and calculate its mass as a stable
state with respect to strong interactions. The approach adopted here 
will be to work with a general, completely antisymmetric wave function 
and determine dynamically which flavor-spin-color-orbital
(TSCL) structures are favored in the constituent quark model (CQM).
The model Hamiltonian used in this work is
fixed by the single hadron spectrum \cite{Paris:2000phd} and the
six-quark (deuteron) properties \cite{Paris:2000bj}. It is given as a sum 
of kinetic energy, pair potentials for OGE and OPE, and confining terms:
\begin{align}
\label{eqn:Hcqm}
\hat{H} &= \sum_{i=1}^5 \left[ m_i - \frac{\hbar^2}{2m_i}\nabla^2_i \right]
+ \sum_{i<j=1}^5 \Bigl[ \frac{1}{4}\st r_{ij} + \hat{v}^g_{ij}(\rvec_{ij})
\nonumber \\
&+ \hat{v}^\pi_{ij}(\rvec_{ij}) \Bigr]
+ V^{FT}(\rvec_1,\ldots,\rvec_5) - V_0(4q\bar{q}).
\end{align}
Here, $m_i$ is the constituent quark mass, $\rvec_{ij}=\rvec_i-\rvec_j$,
and $\st$ is the flux-tube tension.
The two-body operator potentials $\hat{v}^{g(\pi)}_{ij}(\rvec_{ij})$ 
for OGE (OPE) interactions are determined in the non-relativistic reduction
of the tree level amplitude for $qq$ or $q\bar{q}$ scattering. The many-body
confining interaction
\begin{align}
\label{eqn:VFT}
V^{FT}(\bvec{R}) &= \st L(\bvec{R})-\frac{1}{4} \sum_{i<j=1}^5 \st r_{ij}
\end{align}
where $\bvec{R}=(\rvec_1,\ldots,\rvec_5)$ and $L(\bvec{R})$ is the
total length of the flux tubes,
is defined with respect to a two-body contribution (the second term) which
ensures that $V^{FT}(\bvec{R}) \rightarrow 0$ when one of the
quarks is pulled far from the others. The locations of the three $Y$ 
junctions, denoted by $\otimes$ in the diagrams in Fig.(\ref{fig:FTtopo})
are determined by minimizing the total length of the flux tubes given the 
positions $\bvec{R}$ of the quarks and topology of the tubes. Possible 
flux tube topologies for the $4q\bar{q}$ states consistent with gauge 
invariance are shown in Figs.(\ref{fig:FTtopo}). They correspond to overall
singlet states of the form
\begin{align}
\label{eqn:FTbasis}
\frac{1}{2\sqrt{6}}\epsilon_{\alpha\beta\gamma}
\epsilon_{\beta\mu\nu}\epsilon_{\gamma\rho\sigma}
c_1^\mu c_i^\nu c_j^\rho c_k^\sigma \bar{c}_5^\alpha,
\end{align}
where $ijk=234$, $432$, or $324$. Off-diagonal matrix elements of operators
with color dependence (like those in $\hat{v}^g(\rvec_{ij})$) of these
orthonormal states are accompanied by a factor which models the dynamics 
of the gauge fields:
\begin{align}
\label{eqn:gft}
\zeta(r_{il})=e^{-\gft^2 r_{il}^2},
\end{align}
where $\gft^{-1}$ is the range over which flux tubes can be exchanged
(roughly the width of the flux tube) and $r_{il}$ is the distance between 
the exchanged quarks \cite{Paris:2000bj}. We use $\gft=0$, which has been
determined to minimize the energy. The two-body confining interaction
$\frac{1}{4}\st r_{ij}$ of Eq.(\ref{eqn:Hcqm})
is accounted for in VMC non-perturbatively while the $V^{FT}$, typically 
$\sim 15\%$ of the total confinement contribution, is evaluated 
perturbatively. The constant $V_0(4q\bar{q})$ in Eq.(\ref{eqn:Hcqm}), 
extrapolated from the single hadron spectrum, will be discussed below.

The state vector for the five quarks, including two--body 
isospin, spin, color and spatial correlations is given by
\begin{align}
\label{eqn:Psi5}
\ket{\Psi_5} = \mathcal{S}\prod_{i<j}\Fij\ket{\Phi_5}
\end{align}
where the symmetrized product of two--body correlation
operators \Fij\ acts on the uncorrelated TSCL state:
\begin{align}
\label{eqn:Phi5}
\ket{\Phi_5}
&= \frac{1}{\sqrt{3}} \sum_{c=R,G,B} \nonumber \\
&\times\left[
\ket{(TSCL)_{[1^4]};T_z,S_z,c,M_L}\otimes\ket{\bar{s};s_z,\bar{c},m_\ell}
\right]_{J,J_z}.
\end{align}
Here we have four light quarks, $uudd$ (or $4q$) described by the state
$\ket{(TSCL)_{[1^4]};T_z,S_z,c,M_L}$ with isospin, $T,T_z$,
spin $S,S_z$, color $C,c$, and orbital angular momentum $L,M_L$
coupled to the completely antisymmetric state, denoted $(TSCL)_{[1^4]}$. 
The flavor is assumed to be $T=0$ and total spin of $4q$
can takes the values $S=0,1,2$. We neglect
the states with $S=2$ since they are expected to have higher interaction
energies than either $S=0,1$. Color of $4q$ is $C=\mathbf{3}$ since we 
require the state of $4q\bar{q}$ to be a color singlet. 
The antiquark is an $\bar{s}$ with
spin $\shalf$, spin projection $s_z$, color $\bar{c}$ and orbital angular 
momentum $\ell,m_\ell$. The square brackets $[\hphantom{\Phi}]_{J,J_z}$ 
indicate coupling of the $4q$ and $\bar{q}$ to total angular 
momentum $J,J_z$.  We will consider states of either parity, $\pi=\pm 1$
in this work. The state in Eq.(\ref{eqn:Psi5}) is translationally 
invariant and requires no correction due to center-of-mass motion.
\begin{table}[t]
\begin{center}
\begin{tabular}{c@{$=$}c@{$\rightarrow$}c}
\vspace{.2cm}
$\ket{1;1^-}$ &
\,${\tiny\yng(2,2)}_T\times{\tiny\yng(3,1)}_S\times 
{\tiny\yng(2,1,1)}_C\times{\tiny\yng(4)}_L$ &
\,${\tiny\yng(3,1)}_{TS}\times{\tiny\yng(2,1,1)}_{CL}$ \\
\vspace{.2cm}
$\ket{1;1^+}$ &
${\tiny\yng(2,2)}_T\times{\tiny\yng(2,2)}_S\times 
{\tiny\yng(2,1,1)}_C\times{\tiny\yng(3,1)}_L$ &
\,${\tiny\yng(4)}_{TS}\times{\tiny\yng(1,1,1,1)}_{CL}$ \\
\vspace{.2cm}
$\ket{2;1^+}$ &
${\tiny\yng(2,2)}_T\times{\tiny\yng(2,2)}_S\times 
{\tiny\yng(2,1,1)}_C\times{\tiny\yng(3,1)}_L$ &
${\tiny\yng(2,2)}_{TS}\times{\tiny\yng(2,2)}_{CL}$ \\
\vspace{.2cm}
$\ket{3;(0,1)^+}$ &
\,${\tiny\yng(2,2)}_T\times{\tiny\yng(3,1)}_S\times 
{\tiny\yng(2,1,1)}_C\times{\tiny\yng(3,1)}_L$ &
\,${\tiny\yng(3,1)}_{TS}\times{\tiny\yng(2,1,1)}_{CL}$ \\
$\ket{4;(0,1)^+}$ &
\,${\tiny\yng(2,2)}_T\times{\tiny\yng(3,1)}_S\times 
{\tiny\yng(2,1,1)}_C\times{\tiny\yng(3,1)}_L$ &
\,${\tiny\yng(2,1,1)}_{TS}\times{\tiny\yng(3,1)}_{CL}$
\end{tabular}
\end{center}
\caption{\label{tab:symmgroup}
Inner products on $S_4$ of spin, isospin, color, and orbital angular
momentum states of $4q$. States are labeled by an index $n$ which
differentiates the symmetry type for a given parity, 
the angular momentum $J_4$ of the four quarks, and the
parity of the state when the strange antiquark is coupled $\pi=-\pi_4$,
$\ket{n;J_4^{\pi}}$. All the states are to be coupled to overall complete
antisymmetry. States $n=3,4$ may be coupled to angular momentum 
$J_4=0,1,2$. $J_4=2$ is ignored in this work.}
\end{table}

Consider the fully (pair) correlated state vector $\ket{\Psi_5}$ in 
a five-particle basis of spin, isospin, and color. We have
\begin{align}
\label{eqn:Psi5comp}
\ket{\Psi_5(\bvec{R})} = \sum_{n_s=1}^{2^5}\sum_{n_t=1}^2\sum_{n_p=1}^3
\psi_{n_s,n_t,n_p}(\bvec{R}) \ket{n_s,n_t,n_p}
\end{align}
where the sums are over the $2^5=64$ spin states $n_s$, the $T=0$ isospin 
space is spanned by two states, $n_t=1,2$, and color states 
$n_p=1,2,3$. Tensor interactions in the pair potentials 
$\hat{v}^{g(\pi)}(\rvec_{ij})$ populate all of the spin states. In the
color (isospin) sector we may work in a reduced basis since the
interactions are mediated by isovector (color octet) bosons.
The color states of $4q$ span the $C=\mathbf{3}$ irreducible
representation (irrep) of $SU(3)$ and 
the $(3,1)$ irrep of the symmetric group on four objects, $S_4$. 
These will be combined with appropriate states in isospin, spin,
and orbital angular momentum to yield states of complete antisymmetry.
An orthogonal transformation relates these states
to those of Eq.(\ref{eqn:FTbasis}).

The OGE and OPE pair potentials, the confining flux tube potential, and 
the two-body correlation operators, $\Fij$ derived from these 
all depend, variously, on the state $\ket{n_s,n_t,n_p}$. Matrix elements
of potential pair operators appearing in the Hamiltonian 
in $\hat{v}^{g(\pi)}(\rvec_{ij})$
are evaluated in Monte Carlo
as described in Ref.\cite{Paris:2000bj}. The state of the gauge 
field depends on the color configuration of the quarks. Three flux tube 
topologies are displayed in Fig.(\ref{fig:FTtopo}). In the adiabatic
approximation, the locations of the $Y$ junctions are
determined by the minimization of the flux tube length.

The states in Table \ref{tab:symmgroup} are obtained by
coupling the flavor and spin states into irreps
of $S_4$, $(\alpha)_{TS}$, where $T=0$, $S=0,1$ and $(\alpha)$ is the
irrep, given by the Young diagram in the Table. 
Then couple color and orbital angular momentum 
sectors $(\beta)_{CL}$, where $C=\mathbf{3}$ and $L=0,1$.  Finally,
the flavor-spin $(TS)$ and color-orbital $(CL)$
are coupled to give an overall singlet 
$(1^4)\subset (\alpha)_{TS}\times(\beta)_{CL}$.
The possible combinations of inner products over spin, isospin, color, and
orbital angular momentum are given schematically in Table \ref{tab:symmgroup}.
There is one negative parity state, $\ket{1;1^-}$ and six positive parity
states, $\ket{n=1,\cdots,4;J_4^{\pi}}$.
The negative parity state $\ket{1;1^-}$ corresponds to the state of 
$4q$ with positive parity, $\pi_4=+1$ with zero orbital angular momentum.
The positive parity states, $\ket{n=1,\ldots,4;J_4^{+}}$ correspond to 
states of $4q$ with negative parity and include one unit of
orbital angular momentum. We have chosen to neglect the state in which
the strange antiquark carries one unit of orbital angular momentum in 
the present study. This is a prejudice based on diquark correlated wave 
functions \cite{Jaffe:2003sg}. Of course, it only contributes to the 
higher energy positive parity states but a more complete study should 
consider this type of wave function.

\begin{table}[t]
\begin{tabular}{c|c@{\hspace{4mm}}c@{\hspace{4mm}}c@{\hspace{4mm}}c@{\hspace{4mm}}c@{\hspace{4mm}}c@{\hspace{4mm}}c}
$n;J_4^\pi$&$1;1^-$	&$1;1^+$&$2;1^+$&$3;0^+$&$3;1^+$&$4;0^+$&$4;1^+$\\
\hline
$M_{\theta^+}$&	2.22	&2.50	&2.57	&2.75	&2.81	&2.83	&2.88	\\
$\exv{T}$ &	1.68	&2.13	&2.02	&2.03	&2.00	&1.92	&1.90	\\
$\exv{V}$ &	0.92	&0.74	&0.93	&1.10	&1.19	&1.29	&1.36
\end{tabular}
\caption{\label{tab:Evmc} Variational energies of the states of
Table \ref{tab:symmgroup} in GeV. The mass is reported as 
$M_{\theta^+}(n;J_4^\pi) =\exv{T} + \exv{V} - 385.5(5)$ MeV. 
Statistical errors for $M_{\theta^+}$ are $\lesssim 5$ MeV.}
\end{table}
Now that we have the uncorrelated states, we consider the correlated state
vector, $\ket{\Psi_5}$ of Eq.(\ref{eqn:Psi5}). The determination 
of the ground state of the many-body
Schr\"{o}dinger equation can be formulated as a constrained variational
problem. We wish to minimize $\exv{H}$ subject to the constraint that
$\bb\Psi_5|\Psi_5\kb$ remains constant. This problem simplifies greatly 
working within the ansatz of Eq.(\ref{eqn:Psi5}) but solution of the 
constrained variational problem now
yields an upper bound on the ground state energy. The ansatz determines
two-body equations for the $\Fij$ which depend on the reduced mass,
flavor, spin, color, and orbital angular momentum of the pair. Pair 
correlation operators for light quark pairs are denoted $\Fqij$, while 
light quark--strange quark pairs are $\Fsij$. For example, in the case
of two light quarks which have total isospin $T$, color $C$,
orbital angular momentum $L$, and choosing spin of the pair $S=0$,
the pair correlation operator satisfies
\begin{align}
\label{eqn:fT0CL}
-\frac{\hbar^2}{2\mu}
&\left[\frac{d^2\hphantom{f}}{dr^2}-\frac{L(L+1)}{r^2}\right]
\Fqij\ket{T,S=0,C} \nonumber \\
&+\left[v_{T0C}(r) - \lambda_{T0C}(r)\right]\Fqij\ket{T,S=0,C}=0.
\end{align}
The pair potentials in Eq.(\ref{eqn:Hcqm}) are projected into $TSC$
channels, 
$ \hat{v}_{ij}^{g(\pi)}(\rvec)\ket{TSC} = v_{TSC}(r)\ket{TSC} $
and the functions
$\lambda_{TSC}(r)$ are generalizations of Euler-Lagrange parameters
arising in the constrained variational problem.
Their long distance form ensures that boundary conditions required of 
the many-body wave function are satisfied \cite{Pandharipande:1981la}. 
They include a 
small number ($\sim\!10$) of variational parameters which are tuned to 
minimize the variational energy. 
In the case $S=1$ the tensor interaction couples partial waves
$L$ to $L\pm 2$ yielding coupled differential equations.
The form of the $\Fqij$ is chosen to be
\begin{align}
\label{eqn:Fqij}
\Fqij &= f^{(qq)}_c(r_{ij})
\left[ 1 + \sum_{p=2}^{12} u^{(qq)}_p(r_{ij}) \Opij \right],\\
\{\Opij\}_{p=1}^{12} = \{ &\openone,\sisj,\tens\}\otimes\{\openone,\titj\}
\otimes\{\openone,\TT\}. \nonumber
\end{align}
Here $u^{(qq)}_p=f^{(qq)}_p/f^{(qq)}_c$ and the $\sum_p$
includes hyperfine, $\sisj$ and tensor, $\tens$ operators and isospin,
$\titj$ and color dependent $\TT$. The \Fsij\ have only confinement and 
one gluon terms.

\begin{figure}[b]
\includegraphics[width=230pt,keepaspectratio,angle=0,clip]{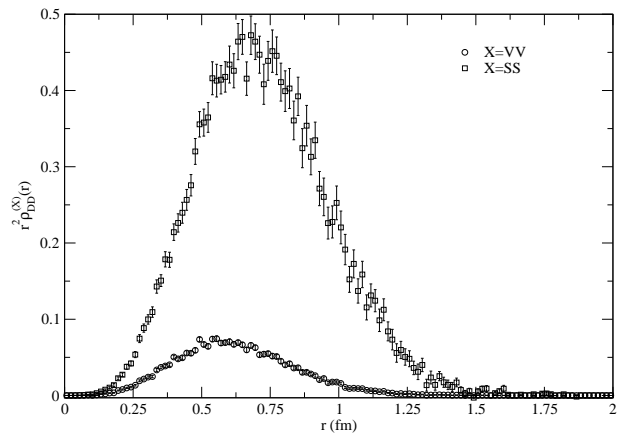}
\caption{\label{fig:rhodd} Radial diquark-diquark pair distributions in the
state $\ket{1;1^+}$.}
\end{figure}
The evaluation of expectation values of the variational wave function
$\ket{\Psi_5}$ are effected via Monte Carlo integration of the 15 dimensional
space $\bvec{R}=(\rvec_1,\ldots,\rvec_5)$. A large number ($\sim 10^5)$
of configurations are sampled from the weight function 
$|\Psi^\dagger_5(\bvec{R})\Psi_5(\bvec{R})|$. Statistical errors are
calculated by block averaging over $\sim\! 500$ configurations
\cite{Lewart:1991}. Results for the variational energies of 
negative and positive parity states are shown in 
Table \ref{tab:Evmc}. The mass of the state is reported as
\begin{align}
\label{eqn:mass}
M_{\theta^+}(n;J_4^{\pi}) = 4m_q + m_s + \exv{T} + \exv{V} - V_0(4q\bar{q})
\end{align}
where $\exv{T}$ and $\exv{V}$ are the variational kinetic and potential
energies, respectively, for each of the states in Table \ref{tab:symmgroup}
and the states are designated by the symmetry index $n$, the total
angular momentum of the four light quarks, and the parity $\pi=-\pi_4$
of the five-quark state $J_4^\pi$. In reality, the
true positive parity ground state is some admixture of the six positive 
parity states. However, the lowest energy state dominates 
since the states $\ket{n;J_4^\pi}$ are orthonormal: applying the 
correlations as in Eq.(\ref{eqn:Psi5}) introduces off-diagonal elements 
which are $\mathcal{O}(u^{(qq)}_p,u^{(qs)}_p)$, quite small. 
States with $n\ne 1$ contribute $\lesssim 10\%$.

The constant $V_0(4q\bar{q})$ is extrapolated from the single baryon
spectrum. In the baryon sector the constant $V_0(3q)$ is fixed by the 
nucleon mass $m_N=939$ MeV to be $1312.6(5)$ MeV where the nucleon mass
and statistical error were calculated in Monte Carlo with $10^5$ 
configurations. We assume that $V_0$ scales linearly with the number
of flux tube ends \cite{Carlson:1991zt,Isgur:1984bm}, 
independent of the flavor of the
associated quark. This gives for the combination
$4m_q+m_s-V_0(4q\bar{q})$ the value $-385.5(5)$ MeV for light quark mass
$m_q=313$ MeV and strange quark mass $m_s=550$ MeV. One might take the
perspective, as some authors have, that the constant should be fitted
to the mass of the ostensible state $\theta^+$. This would give a large
value to the constant $V_0(4q\bar{q})/5=570$ MeV compared with the 
present value $V_0(4q\bar{q})/5=440$ MeV. Of course, our model neglects
effects of chiral symmetry and cannot rule out such an effect. We
note in this context that the present model Hamiltonian does not 
reproduce the correct chiral mass coefficients 
\cite{Jenkins:1991ts,Thomas:1999mu}, a discrepancy that is under current 
study and should be addressed in future work.

Comparison of the CQM with lattice results at large pion mass is 
meaningful since the effects of chiral symmetry are negligible.
We note that the negative parity state $\ket{1;1^-}$ lies below the 
positive parity state $\ket{1;1^+}$, consistent with the lattice data
in Refs.\cite{Mathur:2004jr,Lasscock:2005tt}. This is true despite 
the fact that the positive parity state has much stronger attraction 
from OGE and OPE contributions, attributable to the
high degree of symmetry of the $TS$ state (see Table \ref{tab:Evmc}).
The $\exv{V}$ of the negative parity state is 180 MeV below the lowest
positive parity state. Exciting one unit of orbital angular momentum on 
the other hand raises the kinetic energy significantly, about 450 MeV.

The diquark content of a state gives insight into its short-range
structure. In Figure \ref{fig:rhodd} we plot the diquark-diquark ($DD$)
pair density as a function of radius, for scalar-scalar ($SS$) 
$\rho_{DD}^{(SS)}(r)$ and vector-vector ($VV$) $\rho_{DD}^{(VV)}(r)$ 
diquark configurations for the state $\ket{1;1^+}$. Attraction in
the scalar diquark $T=0$, $S=0$, $C=\overline{\mathbf{3}}$ $SS$ channel
favors this correlation as noted in Ref.\ \cite{Jaffe:2003sg}. It's
about $10$ times larger than the $VV$ correlation but only about
$28\%$ of the total strength of the state. Correlations have a large
effect on this non-perturbative feature of the dynamics: the uncorrelated
state $\Phi_5(1;1^+)$ has equal parts $SS$ and $VV$, constituting
$\third$ of the total strength. Pair correlations induced by the
operators \Fij\ in Eq.(\ref{eqn:Psi5}) therefore significantly
increase the strength of the $SS$ configurations to those of $VV$ but
slightly reduce the relative importance of these correlations in the
overall wave function. The diquark-triquark correlation of 
Ref.\cite{Karliner:2003dt} is orthogonal to the uncorrelated state
$\ket{1;1^+}$ (it appears in states with $n\ne 1$) and is strongly
suppressed in the present calculation.

We have shown that within the non-relativistic flux tube CQM with
OGE and OPE interactions the negative parity state is lower
than the positive parity state by $\sim 280$ MeV when the $\theta^+$
is assumed to be a narrow resonance. Though the higher symmetry of
the lowest lying positive parity state significantly decrease its 
potential energy, its unit excitation of orbital angular momentum 
raises the energy above that of the negative parity
state. While diquark structures are a significant factor in determining
energies of the states, the requisite overall antisymmetry of the 
light quark wave function and pair correlations imply that the majority
of the wave function is described other quark-pair configurations.

\begin{acknowledgments}
The author would like to thank Jozef Dudek, Bob Wiringa, and Ross Young
for helpful discussions. This work was supported by DOE contract 
DE-AC05-84ER40150 Modification No. M175, under which the Southeastern 
Universities Research Association (SURA) operates the Thomas Jefferson 
National Accelerator Facility.
\end{acknowledgments}
\bibliography{q5}
\end{document}